\documentclass[11pt]{article}

\usepackage{amsfonts}
\usepackage{amssymb}
\usepackage{amsmath}
\usepackage{amsthm}
\usepackage{dsfont}
\usepackage{graphics}
\usepackage{graphicx}
\usepackage{epsfig}
\usepackage{mathrsfs}
\usepackage[all]{xy}
\usepackage{color}

\newtheorem{theorem}{Theorem}[section]
\newtheorem{assumption}[theorem]{Assumption}

\newtheorem{corollary}[theorem]{Corollary}
\newtheorem{definition}[theorem]{Definition}
\newtheorem{example}[theorem]{Example}

\newtheorem{lemma}[theorem]{Lemma}

\newtheorem{remark}[theorem]{Remark}

\numberwithin{equation}{section}
\usepackage[colorlinks=true,citecolor=blue]{hyperref}

\def\trieq{\triangleq}

\def\id{\mathbf{1}}
\def\qed{\hbox{ }\hfill {$\Box$}}

\def\proof{\noindent\textbf{Proof. }}

\def\Ept{\mathrm{E}}

\DeclareMathOperator*{\esssup}{ess\,sup}
\DeclareMathOperator*{\essinf}{ess\,inf}
\DeclareMathOperator*{\maxi}{maximize\,}
\DeclareMathOperator*{\mini}{minimize\,}
\DeclareMathOperator*{\argmax}{argmax\,}
\DeclareMathOperator*{\argmin}{argmin\,}

\def\Pbb{\mathbb{P}}

\def\Rbb{\mathbb{R}}
\def\Tbb{\mathbb{T}}

\def\Tbf{\mathbf{T}}

\def\Fc{\mathcal{F}}

\def\Nc{\mathcal{N}}

\def\Xc{\mathcal{X}}

\def\Xsc{\mathscr{X}}

\title{Optimal Investment with Risk Controlled by\\ Weighted Entropic Risk Measures\thanks{Supported by the National Key R\&D Program of China (NO. 2020YFA0712700) and NSFC (NO. 12071146). 
}}
\author{Jianming Xia\thanks{RCSDS, NCMIS, Academy of Mathematics and Systems Science,
Chinese Academy of Sciences, Beijing 100190, China; Email:
xia@amss.ac.cn.}
}

\begin{document}

\maketitle

\begin{abstract}

A risk measure that is consistent with the second-order stochastic dominance and additive for sums of independent random variables can be represented as a weighted entropic risk measure (WERM). The expected utility maximization problem with risk controlled by WERM and a related risk minimization problem are investigated in this paper. The latter  is same to a problem of maximizing a weighted average of constant-absolute-risk-aversion (CARA) certainty equivalents. The solutions of all the optimization problems are explicitly characterized and an iterative method of the solutions is provided.  

\medskip

 \end{abstract}

\section{Introduction}

An important feature in practical investments is managing
market-risk exposure. In the existing literature,  variance, value at risk (VaR), and conditional VaR (CVaR)\footnote{Known also as ``expected shortfall" or
``average VaR" in the literature.} are the most popular three risk measures in studying the problem of portfolio optimization with risk constraints. Among others, Markowitz (1952) investigated the problem of mean-variance efficient portfolio selection, where variance is used as the risk measure; Basak and Shapiro (2001) and Basak
et al. (2006) investigated the problem of expected utility maximization with VaR-based risk management; Rockafellar and Uryasev (2000) investigated the problem of portfolio optimization with CVaR-based risk management.  
Extensions of VaR and CVaR, including spectral risk measures and weighted VaR, have arisen and been applied to the
problem of portfolio optimization with risk management; see 
Acerbi (2002), Acerbi and Simonetti (2002), Adam et al. (2008),
Cahuich and Hern\'andez-Hern\'andez (2013),  He et al. (2015), Ding and Xu (2015), and 
Wei (2018, 2021), for examples.

It is a natural requirement that a risk measure $h$ should have the following two properties.
\begin{itemize}
\item \textit{Monotonicity}: If $X$ first-order stochastically dominates $Y$, then $h(X)\le h(Y)$.
\item \textit{Additivity}: If $X$ and $Y$ are independent, then $h(X+Y)=h(X)+h(Y)$. 
\end{itemize}
The financial meaning of monotonicity is clear: the downside risk of a position is reduced if the payoff profile is increased. The financial meaning of additivity is also clear: the risk associated with a portfolio made of independent assets is exactly the sum of the risk associated with each individual asset.
 A risk measure satisfying monotonicity and additivity is called a \textit{monotone additive risk measure}. 
 
Variance, VaR, and CVaR are, however, not monotone additives risk measures (except for the trivial cases of VaR and CVaR with the extremal confidence levels $\alpha=0,1$).  
Variance is additive but not monotone. VaR and CVaR are monotone but not additive.

Goovaerts et al (2004) and Mu et al (2021) showed that a monotone additive risk measure $h$ has a representation\footnote{The monotonicity of Goovaerts et al (2004) is stronger than that of Mu et al (2021). But their representations are same.}
$$h(X)= \int {1\over a}\log\Ept[e^{-a X}]\, m(da),$$
where $m$ is a probability measure on $[-\infty,\infty]$.
Furthermore, Mu et al (2021) also showed that $h$ is consistent with the second-order stochastic dominance (SSD) if and only if the weighting measure $m$ is supported by $[0,\infty]$. In that case,
$$h(X)= \int_{[0,\infty]} {1\over a}\log\Ept[e^{-a X}]\, m(da).$$
It is well known that ${1\over a}\log\Ept[e^{-a X}]$ is the entropic risk measure with parameter $a>0$; see F\"ollmer and Schied (2016), for example. Therefore, an SSD-consistent monotone additive risk measure is a weighted average of entropic risk measures and hence is called a \textit{weighted entropic risk measure} (WERM).

To the best of our knowledge, the literature to date contains no reports 
on the problem of expected utility maximization with risk controlled by WERMs. The present paper is aimed at filling that gap. As initiating work, we study the problem within a complete market. 

The main contributions of the paper are as follows. 

The feasibility problem of the constraints motives us to consider the problem of risk minimization, which is essentially same to the problem of maximizing a weighted average of CARA certainty equivalents. The existence and uniqueness of the optimal solution is established. The optimality condition for the optimal solution $X^*$ is given as
$$
\begin{cases}
  h^\prime(X^*)+\lambda\rho \ge0\quad \text{a.s.}, \\
  h^\prime(X^*)+\lambda\rho=0\quad \text{a.s. on }(X^*>0),
\end{cases}
$$
where $\rho$ is the stochastic discount factor (SDF)\footnote{Known also as the ``pricing kernel" or ``state price density" in the
literature.}, $\lambda$ is the Lagrangian multiplier, $h^\prime(X^*)$ is given by
$$h^\prime(X)=-\int {e^{-aX}\over\Ept[e^{-aX}]}\,m(da),$$
and $-h^\prime(X)$ is the so-called \textit{marginal risk density}.

For the expected utility maximization problem with a WERM constraint, the existence and uniqueness of the optimal solution is also established. The optimality condition for the optimal solution $X^*$ is then
$$
U^\prime(X^*)-\mu h^\prime(X^*)-\lambda\rho=0,
$$
where $U$ is the utility function and $\lambda$ and $\mu$ are the Lagrangian multipliers. 
 
 For the both optimization problems, an iterative method to the optimal solutions is provided.

The rest of this paper is organized as follows. Section \ref{sec:problem_formulationl} formulates the optimization problems. Section \ref{sec:rmp} investigates the risk minimization problem. Section \ref{sec:eu:werm}  investigates the expected utility maximization problem with a WERM constraint. Section \ref{sec:example} provides numerical examples. Appendix \ref{app:werm} collects some properties of WERM.  Appendix \ref{sec:proofs} presents some proofs.

\newpage

\section{Problem Formulation}\label{sec:problem_formulationl}

\subsection{Expected Utility}\label{sec:eu}

Consider a complete probability space $(\Omega, \Fc, \Pbb)$ and an
arbitrage-free market. Assume that the market is complete and has a
unique SDF $\rho$, which is $\Fc$-measurable and satisfies
$\Pbb(\rho>0)=1$ and $\Ept[\rho]<\infty$.

Consider an agent that is an expected utility maximizer and whose
utility function $U$ satisfies the next standing assumption.
\begin{assumption}\label{ass:U}
$U:(0,\infty)\rightarrow(-\infty,\infty)$ is continuously
differentiable, strictly increasing, and strictly concave and satisfies the Inada conditions:
\begin{eqnarray*}
U^\prime(0)\trieq\lim_{x\downarrow 0}U^\prime(x)=\infty,\quad
U^\prime(\infty)\trieq\lim_{x\to\infty}U^\prime(x)=0.
\end{eqnarray*}
\end{assumption}

Following Kramkov and Schachermayer (1999), we introduce the following assumption
on the asymptotic behavior of the elasticity of $U$.
\begin{assumption}\label{ass:AE} 
$\mathrm{AE}(U)\trieq\limsup_{x\to\infty}{xU^\prime(x)\over U(x)}<1$.
\end{assumption}

Let $L^0$ denote all $\Fc$-measurable random
variables and
$$L^0_+\trieq\left\{X\in L^0\,\left|\, \Pbb(X\ge0)=1\right.\right\}.$$
For any initial capital $x>0$, let $\Xsc(x)$ denote all affordable
terminal wealth:
$$\Xsc(x)\trieq\left\{X\in L^0_+\,\left|\,\Ept[\rho X]\le x\right.\right\}.$$
The standard expected utility maximization problem is
\begin{equation}\label{opt:eu}
\maxi_{X\in\Xsc(x)} \Ept[U(X)]
\end{equation}
and its optimal value is denoted by $u_\circ(x)$. 
To avoid unnecessary technical details, we make the following standing assumption.

\begin{assumption}\label{ass:u_finite}
$u_\circ(x)<\infty$ for some $x>0$.
\end{assumption}

\subsection{Weighted Entropic Risk Measure}

A monotone additive risk measure is a functional $h$ from a subspace $\Xsc\subset L^0$ to $[-\infty,\infty]$ that satisfies the following two properties:
\begin{itemize}
\item Monotonicity: If $X\succsim Y$, then $h(X)\le h(Y)$;
\item Additivity: If $X$ and $Y$ are independent, then $h(X+Y)=h(X)+h(Y)$. 
\end{itemize}
Monotone additive risk measures, including WERMs as special cases, are axiomatically characterized  by Goovaerts et al (2004) and Mu et al (2021); the axiomatic characterization of non-monotone additive risk measures can go back to Gerber and Goovaerts (1981). Goovaerts et al (2004) and Mu et al (2021) showed that a monotone additive risk measure $h$ can be represented as
$$h(X)= \int_{[-\infty,\infty]}h_a(X) m(da),\quad X\in \Xsc.$$
Here $m$ is a probability measure on $[-\infty,\infty]$, and, for every $a\in[-\infty,\infty]$ and $X\in \Xsc$, $h_a(X)$ is given by
\begin{equation}\label{eq:ha:X}
h_a(X)=\begin{cases}
-\esssup X, &a=-\infty\\
    -\Ept[X],   & a=0, \\
   {1\over a}\log\Ept[e^{-a X}],   & a\in(-\infty,\infty),\ a\ne0,\\
  -\essinf X, &a=\infty.
\end{cases}
\end{equation}
In Goovaerts et al (2004),  $X\succsim Y$ represents $h_a(X)\le h_a(Y)$ for all $a$. In Mu et al (2021), $X\succsim Y$ 
represents that $X$ first-order stochastically dominates (FSD) $Y$. The monotonicity of Goovaerts et al (2004) is hence stronger than that of Mu et al (2021). But their representations are same.  

Furthermore, Mu et al (2021) also showed that a monotone additive risk measure $h$  is SSD-consistent if and only if its has representation
\begin{equation}\label{eq:ssd:h}
h(X)= \int_{[0,\infty]}h_a(X) m(da),\quad X\in \Xsc,
\end{equation}
where $m$ is a probability measure on $[0,\infty]$. For each $a>0$, $h_a$ is the well known entropic risk measure with parameter $a$; see F\"ollmer and Schied (2016). Therefore, the representation \eqref{eq:ssd:h} is a weighted average of entropic risk measures according to a weighting measure $m$. For this reason, an SSD-consistent monotone additive risk measure is also called a weighted entropic risk measure. Appendix \ref{app:werm} collects some properties of WERM.

We make the following assumption on the weighting measure $m$.

\begin{assumption}\label{ass:m}
There exists some $0<a_0<a_1<\infty$ such that
$m([a_0,a_1])=1$.
\end{assumption}

Under Assumption \ref{ass:m} we have that 
\begin{equation}\label{eq:hX:finite}
h(X)=\int_{[a_0,a_1]}{1\over a}\log\Ept[e^{-aX}] m(da)\in(-\infty,0],\quad X\in L^0_+.
\end{equation}

\subsection{Optimization Problem}

Given an initial capital $x>0$, the problem that we consider is
maximizing $\Ept[U(X)]$ from the terminal wealth $X$ subject to the budget constraint and the risk constraint:
\begin{equation}\label{opt:eu_X}
\begin{split}
\maxi_{X\in\Xsc(x)}&\quad \Ept[U(X)]\\
\mbox{subject to}&\quad h(X)\le \gamma,
\end{split}
\end{equation}
where $\gamma<0$ is a constant. 
Before studying problem \eqref{opt:eu_X}, we consider the feasibility problem of the constraints: is there a feasible solution, i.e.,
is there an $X$ that satisfies both the budget constraint
$\Ept[\rho X]\le x$ and the risk constraint $h(X)\le \gamma$? This motives us to consider the
risk minimization problem as follows:
\begin{align}\label{opt:rmp}
\mini_{X} h(X) \quad
\mbox{subject to } X\in\Xsc(x).
\end{align}
Let $\gamma_1(x)$ denote the minimal value of problem \eqref{opt:rmp}.
The risk minimization problem will be investigated in the next section.

\begin{remark}
The risk minimization problem is obviously same to the following problem
\begin{align}\label{opt:f}
\maxi_{X} f(X)\quad
\mathrm{subject\ to\ } X\in\Xsc(x),
\end{align}
where 
$$f(X)\trieq -h(X)=-\int_{[a_0,a_1]}{1\over a}\log\Ept[e^{-aX}] m(da)=-\int_{[a_0,a_1]}h_a(X)m(da)$$
is a monotone utility that is additive for the sum of independent random variables. 
The utility $f(X)$ is a weighted average of CARA certainty equivalents since each $-h_a(X)$ is the certainty equivalent of $X$ for CARA utility function $-e^{-ax}$.  As pointed out by Mu et al (2021, Section 4.1), if a monotone preference over monetary gambles is not affected by independent background risks, then it can be represented by a weighted average of CARA certainty equivalents.  Therefore, the next section itself is interesting if the optimization problem for such a preference is concerned.
\end{remark}

\section{Risk Minimization Problem}\label{sec:rmp}

Recall that $\gamma_1(x)$ is the minimal value of problem \eqref{opt:rmp}. Obviously, $\gamma_1$ is decreasing\footnote{Herein, ``increasing" means
``non-decreasing" and ``decreasing" means ``non-increasing."} and convex on $(0,\infty)$. The following theorem establishes the existence and uniqueness of the solution of the risk minimization problem.

\begin{theorem}\label{thm:rmp:unique} 
Under Assumption \ref{ass:m},
problem \eqref{opt:rmp} has a unique solution for every $x>0$. 
\end{theorem}

\proof See Appendix \ref{sec: proof: thm:rmp:unique}. \qed

\medskip

\subsection{Optimality Condition}

Problem \eqref{opt:rmp} can be solved by using Lagrangian multipliers. We begin with the next lemma.

\begin{lemma}\label{lma:opt:rmp:mu0} 
Under Assumption \ref{ass:m}, 
$X^*\in L^0_+$ solves problem \eqref{opt:rmp} if and only if there exists some $\lambda^*>0$ such that
    \begin{align}\label{opt:rmp:mu*}
    \begin{cases}
      X^*\in\displaystyle\argmin_{X\in L^0_+}h(X)+\lambda^*\Ept[\rho X], \\
      \Ept[\rho X^*]=x.
\end{cases}
\end{align} 
If that is the case, then $-\lambda^*\in\partial \gamma_1(x)$, where
$\partial \gamma_1$ denotes the subdifferential of $\gamma_1$.
\end{lemma}

\proof See Appendix \ref{app:proof:lma:opt:rmp:mu0}. \qed

\medskip

Now we consider, for every $\lambda>0$, the problem 
  \begin{align}\label{opt:rmp:mu}
      \mini_{X\in L^0_+} h(X)+\lambda\Ept[\rho X].
\end{align}
The minimal value of problem \eqref{opt:rmp:mu} is denoted by $c_1(\lambda)$, $\lambda>0$. 

Obviously, $c_1$ is concave, increasing, and upper semi-continuous (and hence right-continuous) on $(0,\infty)$.  Let 
$$\lambda_0\trieq\sup\{\lambda\in(0,\infty)\,|\, c_2(\lambda)=-\infty\}.$$
It is easy to see that $\lambda_0\ge{1\over\Ept[\rho]}$ since if $\lambda<{1\over\Ept[\rho]}$ then 
$$c_1(\lambda)\le h(n)+\lambda\Ept[\rho n]=n(\lambda\Ept[\rho]-1)\to-\infty\quad\mbox{as }n\to\infty.$$ 
Therefore, 
$$\left({1\over\Ept[\rho]},\infty\right)\subseteq(\lambda_0,\infty)\subseteq(c_2>-\infty)\subseteq[\lambda_0,\infty).$$

\begin{remark}\label{rmk:emp:lambda}
Under Assumption \ref{ass:m}, similarly to Theorem \ref{thm:rmp:unique}, we can show that problem \eqref{opt:rmp:mu} has a unique solution $X^{*\lambda}$ for each $\lambda\in(c_2>-\infty)$.
Let 
\begin{align}\label{eq:Xc:mu}
\Xc(\lambda)=\Ept[\rho X^{*\lambda}],\quad \lambda\in (c_2>-\infty).
\end{align}
Then Lemma \ref{lma:opt:rmp:mu0} implies that
$\Xc$ is decreasing on $(c_2>-\infty)$. Moreover, a combination of Theorem \ref{thm:rmp:unique} and Lemma \ref{lma:opt:rmp:mu0} implies that, for any $x>0$, $\Xc(\lambda^*)=x$ for some $\lambda^*\in(c_2>-\infty)$.
Therefore, $\Xc$ is continuous on $(c_2>-\infty)$, 
$$\lim_{\lambda\downarrow \lambda_0}\Xc(\lambda)=\infty\quad\mbox{and}\quad 
\lim_{\lambda\uparrow\infty}\Xc(\lambda)=0.$$
The monotonicity and continuity of $\Xc$ make it easy to search for the desired Lagrangian multiplier $\lambda^*$ for any given budget level $x>0$ by solving the equation $\Xc(\lambda)=x$. 
\end{remark}

The optimal solution is closely related to a transformation $h^\prime$ of $h$, which is given by
\begin{equation}\label{eq:hprime}
h^\prime(X)=-\int_{[a_0,a_1]}{e^{-aX}\over\Ept[e^{-aX}]}\,m(da),\quad X\in L^0.
\end{equation}
We now give an economic interpretation of \eqref{eq:hprime} by some heuristic calculations.
Firstly, for any given state $\omega_0\in\Omega$, 
$${\partial \Ept[e^{-aX}]\over\partial X(\omega_0)}={\partial\int e^{-aX(\omega)}\Pbb(d\omega)\over\partial X(\omega_0)}={\partial  e^{-aX(\omega_0)}\Pbb(d\omega_0)\over\partial X(\omega_0)}=-a e^{-aX(\omega_0)}\Pbb(d\omega_0).$$
The marginal risk of losing money in state $\omega$ is then
\begin{align*}
-{\partial h(X)\over\partial X(\omega)}=
&-\int_{[a_0,a_1]}{1\over a} {\partial\log\Ept[e^{-aX}]\over\partial X(\omega)} m(da)\\
=&-{\int_{[a_0,a_1]}{1\over a \Ept[e^{-aX}]}{\partial \Ept[e^{-aX}]\over\partial X(\omega)}}m(da)\\
=&\int_{[a_0,a_1]}{e^{-aX(\omega)}\over\Ept[e^{-aX}]}m(da)\Pbb(d\omega),\quad \omega\in\Omega.
\end{align*}
The marginal risk density of state $\omega$ is then 
$$-{\partial h(X)/\partial X(\omega)\over\Pbb(d\omega)}=\int_{[a_0,a_1]}{e^{-aX(\omega)}\over\Ept[e^{-aX}]}m(da),\quad \omega\in\Omega.$$
Therefore, $-h^\prime(X)$ is the \textit{marginal risk density} of $h$.  Moreover, Theorem  \ref{thm:h:direct:derivative} in the appendix shows that $h^\prime$ is something like the G\^ateaux derivative of $h$.

The next theorem explicitly characterizes the solution of problem \eqref{opt:rmp:mu} in terms of the marginal risk density.   

\begin{theorem}\label{thm:rmp}
Let $m$ satisfy Assumption \ref{ass:m} and $\lambda\in (c_2>-\infty)$. Then $X^*\in L^0_+$ solves problem \eqref{opt:rmp:mu} if and only if it satisfies the following condition 
\begin{equation}\label{eq:solution:rmp}
\begin{cases}
h^\prime(X^*)+\lambda\rho\ge0\quad \text{a.s.}, \\
h^\prime(X^*)+\lambda\rho=0\quad \text{a.s. on }(X^*>0).
\end{cases}
\end{equation}
\end{theorem}

\proof See Appendix \ref{app:rmp}.\qed

\medskip

The optimality condition \eqref{eq:solution:rmp} is an explicit  trade-off between the risk and the price by comparing the marginal risk density $-h^\prime(X^*)$ and the state price density $\rho$.

A combination of Theorems \ref{thm:rmp:unique}  and  \ref{thm:rmp} and Lemma \ref{lma:opt:rmp:mu0}  yields the following corollary.

\begin{corollary} \label{cor:emp:x}
Let $m$ satisfy Assumption \ref{ass:m}. For every $x>0$, problem \eqref{opt:rmp} has a unique solution $X^*$. Moreover, any $X^*\in L^0_+$ is the solution if and only if it satisfies condition \eqref{eq:solution:rmp} for some some $\lambda>0$ with $\Ept[\rho X^*]=x$.
\end{corollary}

In the next part of this subsection, we write the optimality condition \eqref{eq:solution:rmp} in another form. 
Let functions $\varphi$ and $\psi$ be given by
\begin{align}\label{eq:varphi:psi}
\begin{cases}
\varphi(a)=\Ept[e^{-aX^*}],\quad a\in[a_0,a_1],\\
\psi(y)=\displaystyle\int_{[a_0,a_1]}{e^{-ay}\over\varphi(a)}m(da),\quad y\in\Rbb.
\end{cases}
\end{align}
It is easy to see that both $\varphi$ and $\psi$ are continuous and strictly decreasing with
$$\lim_{y\downarrow-\infty}\psi(y)=\infty\quad\mbox{and}\quad\lim_{y\uparrow\infty}\psi(y)=0.$$

\begin{lemma}\label{lma:X*g}
The optimality condition \eqref{eq:solution:rmp} is equivalent to 
\begin{equation}\label{eq:X*:psi+}
X^*=\left(\psi^{-1}\left({\lambda\rho}\right)\right)^+,
\end{equation}
where $\psi$ is given by \eqref{eq:varphi:psi}.
\end{lemma}

\proof  Recall that $\psi$ is strictly decreasing and $h^\prime(X^*)=-\psi(X^*)$. Then condition \eqref{eq:solution:rmp} reads
\begin{equation*}
\begin{cases}
\lambda \rho-\psi(0)\ge0\quad \text{a.s. on }(X^*=0), \\
  \lambda\rho-\psi(0)<\lambda\rho-\psi(X^*)=0\quad \text{a.s. on }(X^*>0),
\end{cases}
\end{equation*} 
which is equivalent to
\begin{equation}\label{eq:solution:rmp:gX*}
\begin{cases}
  X^*=0    & \text{if }\lambda\rho\ge\psi(0), \\
 \lambda\rho=\psi(X^*)    & \text{if } \lambda\rho<\psi(0).
\end{cases}
\end{equation} 
Finally, we can rewrite \eqref{eq:solution:rmp:gX*} in form of \eqref{eq:X*:psi+}.\qed

\medskip

In view of \eqref{eq:X*:psi+}, $X^*$ is determined by $\psi$, which should be further jointly determined with $\varphi$.
We now derive the equations for $(\varphi,\psi)$ by plugging \eqref{eq:X*:psi+} into \eqref{eq:varphi:psi}.
By \eqref{eq:X*:psi+}, the probability distribution function $F_{X^*}$ of $X^*$ satisfies that
$$1-F_{X^*}(y)=\Pbb(X^*>y)=\Pbb(\lambda\rho<\psi(y)),\quad y\ge0.$$
Recalling the following formula, see Feller (1971, Section 13.2) for example,
\begin{equation*}
\varphi(a)=\Ept[e^{-aX^*}]=1-a\int_0^\infty e^{-ay} (1-F_{X^*}(y)) dy,
\end{equation*}
we have that
\begin{equation}\label{eq:X*:laplace}
 \varphi(a)=1-a\displaystyle\int_0^\infty e^{-a y} F_\rho\left({\psi(y)\over\lambda}-\right)dy=1-a\displaystyle\int_0^\infty e^{-a y} F_\rho\left({\psi(y)\over\lambda}\right)dy
 \end{equation}
since $\psi$ is strictly decreasing and hence
$$F_\rho\left({\psi(y)\over\lambda}\right)\ne F_\rho\left({\psi(y)\over\lambda}-\right) \quad\mbox{for at most countable }y.$$ 
Therefore, we get the following system of integral equations for $(\varphi,\psi)$:
\begin{align}\label{eqs:varphi:psi:F}
\begin{cases}
\varphi(a)=1-a \displaystyle\int_0^\infty e^{-ay}F_\rho\left({\psi(y)\over\lambda}\right)dy,\quad a\in[a_0,a_1],\\
\psi(y)=\displaystyle\int_{[a_0,a_1]}{e^{-ay}\over\varphi(a)}m(da),\quad y\in\Rbb.
\end{cases}
\end{align}

\subsection{Iterative Method of the Optimal Solution}\label{sec:iteration:rmp}

Now we provide an iterative method to solve problem \eqref{opt:rmp:mu}.
For any given $X\in L^0_+$, let
\begin{align}\label{eq:phiX:psiX}
\begin{cases}
\varphi_X(a)=\Ept[e^{-aX}],\quad a\in[a_0,a_1], \\
 \psi_X(y)=\displaystyle\int_{[a_0,a_1]}{e^{-ay}\over\varphi_X(a)}m(da),\quad y\in\Rbb. \\
\end{cases}
\end{align}
Obviously, both $\varphi_X$ and $\psi_X$ are continuous and strictly decreasing. Moreover, 
\begin{equation}\label{inada:psiX}
\lim_{y\to-\infty}\psi_X(y)=\infty\mbox{ and }\lim_{y\to\infty}\psi_X(y)=0.
\end{equation}

For any given $\lambda>0$, we consider a transform $\Tbf: L^0_+\to L^0_+$ defined as follows. 

\begin{definition}\label{def:T}
For any given $X\in L^0_+$, let
\begin{align*}
 \Tbf X=\left(\psi_X^{-1}\left({\lambda\rho}\right)\right)^+.
\end{align*}
\end{definition}

Obviously, \eqref{eq:X*:psi+} is equivalent to that $X^*$ is a fixed point of $\Tbf$: $X^*=\Tbf X^*$.

\begin{lemma}\label{lma:T:mono}
The transform $\Tbf$ in Definition \ref{def:T} is monotone: for any $X_1,X_2\in L^0_+$, we have that 
$$X_1\ge X_2\mbox{ a.s. } \Rightarrow\ \Tbf X_1\ge \Tbf X_2 \mbox{ a.s.} $$
\end{lemma}
\proof It is easy and standard.\qed

Now we consider a sequence $\{X_n,n\ge0\}\subset L^0_+$ given by
\begin{align}\label{eq:X0Xn}
X_0\equiv0\quad\mbox{and}\quad X_{n+1}=\Tbf X_n\mbox{ for all }n\ge0.
\end{align}
By Lemma \ref{lma:T:mono} and $X_1\ge0=X_0$ a.s., we have that $X_2=\Tbf X_1\ge\Tbf X_0=X_1$ a.s. 
Successively continuing the above argument leads to $X_{n+1}\ge X_n$ a.s. for all $n\ge0$. Then we know that 
\begin{equation}\label{eq:Xn:Xinfty}
X_n\nearrow X_\infty\mbox{ a.s.}
\end{equation}
for some $[0,\infty]$-valued random variable $X_\infty$. 

\begin{theorem}\label{thm:rmp:iteration} 
Let $m$ satisfy Assumption \ref{ass:m} and $\lambda>0$. Then problem \eqref{opt:rmp:mu} has a solution if and only if $\Ept[\rho X_\infty]<\infty$. If that is the case, then $X_\infty$ solves problem \eqref{opt:rmp:mu}. 
\end{theorem}

\proof Assume that $X^*$ solves problem \eqref{opt:rmp:mu}. Then $X^*=\Tbf X^*\ge \Tbf X_0=X_1$ a.s. by Lemma \ref{lma:T:mono}. Similarly, $X^*=\Tbf X^*\ge \Tbf X_1=X_2$ a.s. Successively continuing the above argument leads to
$X^*\ge X_n$ a.s. for all $n\ge0$. Therefore $X^*\ge X_\infty$ a.s. and hence $\Ept[\rho X_\infty]\le\Ept[\rho X^*]<\infty$.

Conversely, assume that $\Ept[\rho X_\infty]<\infty$. 
Now we show that $X_\infty$ solves problem \eqref{opt:rmp:mu}. It suffices to show that $X_\infty=\Tbf X_\infty$.
To this end, for all $n\ge0$ (including the case $n=\infty$), let
\begin{align*}
\begin{cases}
\varphi_n(a)=\Ept[e^{-aX_n}],\quad a\in[a_0,a_1], \\
 \psi_n(y)=\displaystyle\int_{[a_0,a_1]}{e^{-ay}\over\varphi_n(a)}m(da),\quad y\in\Rbb.
\end{cases}
\end{align*}
By \eqref{eq:Xn:Xinfty} and the monotone convergence theorem,
$$\varphi_n\searrow \varphi_\infty\quad\mbox{everywhere on }[a_0,a_1].$$
By $\Ept[\rho X_\infty]<\infty$, there exists some $\varepsilon>0$ such that
$$\varphi_\infty(a)\ge\varepsilon\quad\mbox{for all }a\in[a_0,a_1].$$
Then by the monotone convergence theorem once again,
$$\psi_n\nearrow \psi_\infty\quad\mbox{everywhere on }\Rbb.$$
Every $\psi_n$ (including $\psi_\infty$) is continuous and strictly decreasing on $\Rbb$ with
$$\lim_{y\downarrow-\infty}\psi_n(y)=\infty\quad\mbox{and}\quad\lim_{y\uparrow\infty}\psi_n(y)=0,\quad n\ge1.$$
Then we have that\footnote{Obviously, $\psi^{-1}_1\le\psi^{-1}_2\le\dots\le\psi^{-1}_n\le\dots\le\psi^{-1}_\infty$ on $\Rbb$. Suppose on the contrary of \eqref{lim:psi:n:-1} that $\psi^{-1}_n(z_0)\nearrow y_0<\psi^{-1}_\infty(z_0)$ for some  $z_0\in(0,\infty)$ and $y_0\in\Rbb$. Then 
$\psi_n(y_0)\le\psi_n(\psi^{-1}_n(z_0))=z_0<\psi_\infty(y_0)$, which contradicts $\psi_n\nearrow \psi_\infty$.}
\begin{equation}\label{lim:psi:n:-1}
\psi_n^{-1}\nearrow \psi_\infty^{-1}\quad\mbox{everywhere on }(0,\infty).
\end{equation}
Recall that, for all $n\ge0$,
$$X_{n+1}=\left(\psi_n^{-1}\left({\lambda\rho}\right)\right)^+ \quad \text{a.s.}.$$
By passing to the limit, we get that $X_\infty=\Tbf X_\infty$.
\qed

\section{Expected Utility Maximization Problem}\label{sec:eu:werm}

Now we go back to the expected utility maximization problem \eqref{opt:eu_X}.

By Theorem \ref{thm:rmp:unique},  problem \eqref{opt:eu_X} has a feasible solution if and only if $\gamma\ge\gamma_1(x)$. Moreover, in the case when $\gamma=\gamma_1(x)$,  problem \eqref{opt:eu_X} is solved by the risk minimizer, which has been characterized by Section \ref{sec:rmp}. 
On the other hand, for every $x>0$, let $X_{\circ x}$ solve expected utility maximization problem \eqref{opt:eu} without risk constraint and let 
\begin{equation}\label{eq:gamma2}
\gamma_2(x)\trieq h(X_{\circ x}),\quad x>0.
\end{equation}
Obviously, if $\gamma\ge\gamma_2(x)$, then $X_{\circ x}$ solve problem \eqref{opt:eu_X} as well. Therefore, in the sequel,  we always assume that $\gamma_1(x)<\gamma<\gamma_2(x)$.

We first establish the existence and uniqueness of the solution.

For any fixed $x>0$, let $u_x(\gamma)$ denote the value of problem \eqref{opt:eu_X}.

\begin{theorem}\label{thm:u2_exists}
Under Assumptions \ref{ass:U}--\ref{ass:m}, for any fixed $x>0$ and any $\gamma\in(\gamma_1(x),\gamma_2(x))$, problem \eqref{opt:eu_X} has a
unique optimal solution $X^*$. Moreover, the value function $u_x$ is continuous and strictly concave on $(\gamma_1(x),\gamma_2(x))$.
\end{theorem}

\proof It is standard and similar to Wang and Xia (2021, Theorem 3.4), by combining some arguments in the proofs of Theorem \ref{thm:rmp:unique}. \qed

\subsection{Optimality Condition}

Problem \eqref{opt:eu_X} can be solved by using Lagrangian
multipliers. 
We begin with the following lemma, which can be proved in a similar way to Lemma \ref{lma:opt:rmp:mu0}.

\begin{lemma}\label{lma:duality:mu:gamma}
Under Assumptions \ref{ass:U}--\ref{ass:m}, for any fixed $x>0$ and any $\gamma\in(\gamma_1(x),\gamma_2(x))$,  $X^*\in\Xsc(x)$ solves problem \eqref{opt:eu_X} if and only if there exists some
$\mu^*>0$ such that 
\begin{equation*}
\begin{cases}
X^*\in\displaystyle\argmax_{X\in \Xsc(x)} \Ept[U(X)]-\mu^*h(X),\\
h(X^*)=\gamma.
\end{cases}
\end{equation*}
If that is the case, then $\mu^*\in\partial u_x(\gamma)$, where
$\partial u_x$ denotes the superdifferential of the value function $u_x$.
\end{lemma}

\medskip

Now we consider,  for every $\mu>0$,  the problem
\begin{equation}\label{opt:eu:mu0}
\maxi_{X\in \Xsc(x)} \Ept[U(X)]-\mu h(X).
\end{equation}

\begin{remark}\label{rmk:vlambda} 
Under Assumptions \ref{ass:U}--\ref{ass:m}, 
similarly to Theorem \ref{thm:u2_exists}, we can show that problem \eqref{opt:eu:mu0} has a unique solution $X^{*\mu}$ for each $\mu>0$.
Let 
\begin{align}\label{eq:Gamma:mu}
\Gamma(\mu)=h(X^{*\mu}),\quad \mu\in(0,\infty).
\end{align}
Then Lemma \ref{lma:duality:mu:gamma} implies that
$\Gamma$ is decreasing on $(0,\infty)$. Moreover, a combination of Theorem \ref{thm:u2_exists} and Lemma \ref{lma:duality:mu:gamma} implies that, for any $\gamma\in(\gamma_1(x),\gamma_2(x))$, $\Gamma(\mu^*)=\gamma$ for some $\mu^*\in(0,\infty)$.
Therefore, $\Gamma$ is continuous on $(0,\infty)$, 
$$\lim_{\mu\downarrow0}\Gamma(\mu)=\gamma_2(x)\quad \mbox{and}\quad \lim_{\mu\uparrow\infty}\Gamma(\mu)=\gamma_1(x).$$
This allows us to search for the desired Lagrangian multiplier $\mu^*$ for any given risk level $\gamma\in(\gamma_1(x),\gamma_2(x))$ by solving the equation $\Gamma(\mu)=\gamma$. 
\end{remark}

Now we consider problem \eqref{opt:eu:mu0} for any fixed $\mu>0$, which reads
\begin{equation}\label{opt:eu:mu:x}
\begin{split}
\maxi_{X\in L^0_+}\quad&\Ept[U(X)]-\mu h(X)\\
\mbox{subject to}\quad& \Ept[\rho X]\le x.
\end{split}
\end{equation}
The value of problem \eqref{opt:eu:mu:x} is denoted by $v_\mu(x)$. It is easy to see that $v_\mu$ is increasing and concave on $(0,\infty)$.  
 
Similarly to Lemma \ref{lma:opt:rmp:mu0}, we have the next lemma.

\begin{lemma}\label{lma:opt:eu:mu*} 
Under Assumptions \ref{ass:U}--\ref{ass:m}, for any fixed $x>0$ and $\mu>0$, 
 $X^*\in L^0_+$ solves problem \eqref{opt:eu:mu:x} if and only if there exist some $\lambda^*>0$ such that
    \begin{align}\label{opt:eu:mu*}
    \begin{cases}
      X^*\in\displaystyle\argmax_{X\in L^0_+}\Ept[U(X)]-\mu h(X)-\lambda^*\Ept[\rho X], \\
       \Ept[\rho X^*]=x.
\end{cases}
\end{align} 
If that is the case, then $\lambda^*\in\partial v_\mu(x)$, where
$\partial v_\mu$ denotes the superdifferential of $v_\mu$.
\end{lemma}

From now on, we focus on studying the problem  
\begin{equation}\label{opt:euwe:lm}
\maxi_{X\in L^0_+}\Ept[U(X)]-\mu h(X)-\lambda\Ept[\rho X],
\end{equation}
where $\lambda>0$ and $\mu>0$. 
The value of problem \eqref{opt:euwe:lm} is denoted by $w(\mu,\lambda)$.

It is easy to see that,  for any given $\mu>0$,
 $w_\mu\trieq w(\mu,\cdot)$ is convex, decreasing, and lower semi-continuous (and hence right-continuous) on $(0,\infty)$.
Let 
$$\lambda_1(\mu)\trieq\sup\{\lambda\in(0,\infty)\,|\, w_\mu(\lambda))=\infty\},\quad\mu>0.$$
It is easy to see that $\lambda_1(\mu)\ge{\mu\over\Ept[\rho]}$ since if $\lambda<{\mu\over\Ept[\rho]}$ then 
$$w_\mu(\lambda)\ge U(n)-\mu h(n)-\lambda\Ept[\rho n]=U(n)+n(\mu-\lambda\Ept[\rho])\to\infty\quad\mbox{as }n\to\infty.$$ 
Therefore, 
$$\left({\mu\over\Ept[\rho]},\infty\right)\subseteq(\lambda_1(\mu),\infty)\subseteq(w_\mu<\infty)\subseteq[\lambda_1(\mu),\infty).$$

Similarly to Remark \ref{rmk:vlambda}, we have the next remark.

\begin{remark}\label{rmk:v2:mu} 
Under Assumptions \ref{ass:U}--\ref{ass:m},  let $\mu>0$ be given. For every $\lambda\in(w_\mu<\infty)$,
we can show that problem \eqref{opt:euwe:lm} has a unique solution $X^{*\mu\lambda}$.
Let 
\begin{align}\label{eq:Xclambda}
\Xc_\mu(\lambda)=\Ept[\rho X^{*\mu\lambda}],\quad \lambda\in(w_\mu<\infty).
\end{align}
Then $\Xc_\mu$ is decreasing and continuous on $(\lambda_1(\mu),\infty)$ with 
$$\lim_{\lambda\downarrow \lambda_1(\mu)}\Xc_\mu(\lambda)=\infty\quad\mbox{and}\quad\lim_{\lambda\uparrow\infty}\Xc_\mu(\lambda)=0.$$ 
This allows us to search for the desired Lagrangian multiplier $\lambda^*$ for any given budget level $x>0$ by solving the equation $\Xc_\mu(\lambda)=x$. 
\end{remark}

The next theorem explicitly characterizes the solution of problem \eqref{opt:euwe:lm}.

\begin{theorem}\label{thm:eu:X}
Under Assumptions \ref{ass:U}--\ref{ass:m}, for any $\mu>0$, $\lambda\in(w_\mu<\infty)$, and $X^*\in L^0_+$, 
We have the following assertions.
\begin{description}
\item[(a)] If $X^*$ solves problem \eqref{opt:euwe:lm}, then $X^*>0$ a.s.

\item[(b)] $X^*$ solves problem \eqref{opt:euwe:lm} if and only if it satisfies condition 
\begin{eqnarray}\label{eq:eu:opt:con}
U^\prime(X^*)-\mu h^\prime(X^*)-\lambda\rho=0\mbox{ a.s.}
\end{eqnarray} 
\end{description}
\end{theorem}

\proof See Appendix \ref{app:proof:thm:eu:X*}.\qed

\medskip

The optimality condition \eqref{eq:eu:opt:con} is an explicit trade-off among the utility, the price, and risk by comparing the marginal utility $U^\prime$,  state price density $\rho$, and the marginal risk density $-h^\prime(X^*)$.

A combination of Theorems \ref{thm:u2_exists} and \ref{thm:eu:X} and Lemmas \ref{lma:duality:mu:gamma} and \ref{lma:opt:eu:mu*} yields the following corollary.

\begin{corollary} 
Under Assumptions \ref{ass:U}--\ref{ass:m}, for every $x>0$ and $\gamma\in(\gamma_1(x),\gamma_2(x))$, problem \eqref{opt:eu_X} has a unique solution $X^*$. Moreover, any $X^*\in L^0_+$ is the solution if and only if it satisfies condition \eqref{eq:eu:opt:con} for some some $\lambda>0$ and $\mu>0$ with $\Ept[\rho X^*]=x$ and $h(X^*)=\gamma$.
\end{corollary}

Let functions $\phi$ be given by  
\begin{equation}\label{eq:phi}
\phi(y)=U^\prime(y)+\mu\psi(y), \quad y\in(0,1),
\end{equation}
where function $\psi$ is defined by \eqref{eq:varphi:psi}. 
Then the optimality condition \eqref{eq:eu:opt:con} reads
\begin{equation}\label{eq:U:psi:X}
\phi(X^*)=\lambda\rho.
\end{equation}
In view of \eqref{eq:U:psi:X}, $X^*$ is determined by $\phi$, which should be further jointly determined with $\varphi$.
Now we derive the equations for $(\varphi,\phi)$. The distribution function $F_{X^*}$ satisfies that
\begin{equation}\label{eq:pdf:X}
1-F_{X^*}(y)=\Pbb(X^*>y)=\Pbb(\phi(X^*)<\phi(y))=\Pbb(\lambda\rho< \phi(y)).
\end{equation}
Therefore, similarly to \eqref{eq:X*:laplace}, we get the following system of integral equations for $(\varphi,\phi)$.  
\begin{equation}\label{eq:integral:varphi:phi}
\begin{cases}
\varphi(a)=1-a\displaystyle\int_0^\infty e^{-a y}F_\rho\left({\phi(y)\over\lambda}\right)dy,\quad a\in[a_0,a_1],\\
\phi(y)=U^\prime(y)+\mu\displaystyle\int_{[a_0,a_1]}{e^{-ay}\over\varphi(a)}m(da),\quad y>0.
\end{cases}
\end{equation}

\subsection{Iterative Method of the Optimal Solution}\label{sec:iteration:eu}

Now we provide an iterative method to solve problem \eqref{opt:euwe:lm}.

Recall that $U$ satisfies the Inada condition and function $\psi_X$ is given by \eqref{eq:phiX:psiX}. For any given $X\in L^0_+$, by \eqref{inada:psiX}, equation \begin{equation}\label{eq:up:Y}
U^\prime(Y)+\mu\psi_X(Y)=\lambda
\end{equation}
has a unique solution $Y\in L^0_+$. Obviously,  $Y>0$ a.s.
We consider a transform $\Tbb: L^0_+\to L^0_+$ defined as follows. 

\begin{definition}\label{def:Tbb} For any given $X\in L^0_+$, let $Y$ be determined by \eqref{eq:up:Y}.
Then $\Tbb X\trieq Y$.
\end{definition}

Obviously, \eqref{eq:U:psi:X} is equivalent to $X^*=\Tbb X^*$.

Similarly to Lemma \ref{lma:T:mono}, transform $\Tbb$ in Definition \ref{def:Tbb} is monotone.
Now we consider a sequence $\{Y_n,n\ge0\}\subset L^0_+$ given by
\begin{align}\label{eq:Y0Yn}
Y_0\equiv0\quad\mbox{and}\quad Y_{n+1}=\Tbb Y_n\mbox{ for all }n\ge0.
\end{align}
By the monotonicity of $\Tbb$ and the same argument as in Section \ref{sec:iteration:rmp}, we can see that 
$$Y_n\nearrow Y_\infty\mbox{ a.s.}$$
for some $(0,\infty]$-valued random variable $Y_\infty$.  Similarly to Theorem \ref{thm:rmp:iteration}, we have the next theorem.

\begin{theorem}\label{thm:eu:iteration}
 Under Assumptions \ref{ass:U}--\ref{ass:m},  problem \eqref{opt:euwe:lm} has a solution $X^*$ with $\Ept[\rho X^*]<\infty$  if and only if $\Ept[\rho Y_\infty]<\infty$. If that is the case, then $Y_\infty$ solves problem \eqref{opt:euwe:lm}. 
\end{theorem}

\section{Numerical Example}\label{sec:example}

\subsection{Entropic Risk Measure}

In this subsection we consider a simple but illustrative case when $h$ is an entropic risk measure:
$$h(X)={1\over a}\log\Ept[e^{-aX}],$$ where parameter $a>0$. In this case, 
$$h^\prime(X)=-{e^{-aX}\over\Ept[e^{-aX}]}.$$  

\subsection*{Risk Minimization}

Assume that $X^{*\lambda}$ solves problem \eqref{opt:rmp:mu}.  
Let 
$$c(\lambda)\trieq\Ept[e^{-aX^{*\lambda}}]\quad\mbox{and}\quad c_1(\lambda)\trieq c(\lambda)\lambda.$$ 
Then by  \eqref{eq:varphi:psi},
$$\psi(y)={e^{-ay}\over c(\lambda)}\quad\mbox{and}\quad\psi^{-1}(z)=-{\log c(\lambda)+\log z\over a}.$$
Therefore, by Lemma \ref{lma:X*g},
$$X^{*\lambda}=\left(\psi^{-1}(\lambda\rho)\right)^+=\left(-{\log c_1(\lambda)+\log \rho\over a}\right)^+.$$
By Remark \ref{rmk:emp:lambda},
$$\Xc(\lambda)=\Ept\left[\rho\left(-{\log c_1(\lambda)+\log \rho\over a}\right)^+\right].$$

Let $c_1(\lambda)$ be determined by 
$$\Ept\left[\rho\left(-{\log c_1(\lambda)+\log \rho\over a}\right)^+\right]=x.$$
Then, by Corollary \ref{cor:emp:x} and Lemma \ref{lma:X*g},  
$$X^*=X^{*\lambda}=\left(-{\log c_1(\lambda)+\log \rho\over a}\right)^+$$ 
solves the risk minimization problem \eqref{opt:rmp}.
Moreover, 
\begin{equation*}
c(\lambda)=\Ept[e^{-aX^{*\lambda}}]=\Ept\left[\left(e^{-a\psi^{-1}(\lambda\rho)}\right)\land1\right]
=\Ept[\left(c(\lambda)\psi(\psi^{-1}(\lambda\rho))\right)\land 1]
=\Ept[\left(c_1(\lambda)\rho\right)\land 1],
\end{equation*}
implying that 
$$\gamma_1(x)={1\over a}\log \Ept[e^{-aX^*}]={1\over a}\log c(\lambda)={1\over a}\log\Ept[\left(c_1(\lambda)\rho\right)\land 1].$$
So we get the closed-form solution for the risk minimization problem in the case when $h$ is an entropic risk measure.

\begin{example} [Log-Normally Distributed SDF]
Assume that the SDF $\rho$ is log-normally distributed:
$\log\rho\sim \Nc (b,\sigma^2)$, where $b\in\Rbb$ and $\sigma>0$ are constants. Let 
$$N(x)\trieq {1\over\sqrt{2\pi}}\int_{-\infty}^x e^{-{z^2\over2}}dz,\quad x\in\Rbb.$$
In this case, we have that
\begin{eqnarray*}
\Xc(\lambda)&=&\Ept\left[\rho\left(-{\log c_1(\lambda)+\log\rho\over a}\right)^+\right]\\
&=&
-{1\over a}\int_{-\infty}^{-{b+\log c_1(\lambda)\over\sigma}}e^{b+\sigma x}(b+\sigma x+\log c_1(\lambda)){1\over\sqrt{2\pi}}e^{-{x^2\over2}}dx\\
&=&-{1\over a}e^{b+{\sigma^2\over2}}\int_{-\infty}^{-{b+\log c_1(\lambda)\over\sigma}}(b+\sigma x+\log c_1(\lambda)){1\over\sqrt{2\pi}}e^{-{(x-\sigma)^2\over2}}dx\\
&=&-{1\over a}e^{b+{\sigma^2\over2}}\left((b+\log c_1(\lambda))N\left(-{b+\log c_1(\lambda)\over\sigma}-\sigma\right)\right.\\
&&\qquad\qquad\qquad\left. +\sigma\int_{-\infty}^{-{b+\log c_1(\lambda)\over\sigma}-\sigma}{1\over\sqrt{2\pi}}(x+\sigma)e^{-{x^2\over2}}dx\right).
\end{eqnarray*}
We can search for the desired $c_1(\lambda)$ by solving the equation $\Xc(\lambda)=x$. Then the risk minimizer 
$$X^*=\left(-{\log c_1(\lambda)+\log\rho\over a}\right)^+.$$ 
Moreover, $\gamma_1(x)={1\over a}\log c(\lambda)$, where
\begin{eqnarray*}
c(\lambda)&=&\Ept[\left(c_1(\lambda)\rho\right)\land 1]\\
&=&c_1(\lambda)\int_{-\infty}^{-{b+\log c_1(\lambda)\over\sigma}}e^{b+\sigma x}{1\over\sqrt{2\pi}}e^{-{x^2\over2}}dx+{1\over c}N\left({b+\log c_1(\lambda)\over\sigma}\right)\\
&=&\lambda e^{b+{\sigma^2\over2}}\int_{-\infty}^{-{b+\log c_1(\lambda)\over\sigma}}{1\over\sqrt{2\pi}}e^{-{(x-\sigma)^2\over2}}dx+{1\over c}N\left({b+\log c_1(\lambda)\over\sigma}\right)\\
&=&\lambda e^{b+{\sigma^2\over2}}N\left(-{b+\log c_1(\lambda)\over\sigma}-\sigma\right)+{1\over c}N\left({b+\log c_1(\lambda)\over\sigma}\right).
\end{eqnarray*}
Figure \ref{fig:gamma12_entropy} displays $\gamma_1$ when $a=2$ and $\log\rho\sim\Nc(-0.5,1)$. (In this case, $\Ept[\rho]=1$.) Particularly, $\gamma_1(1)=-1.2489$.
\end{example}

\subsection*{Expected Utility Maximization}

Recall that $X_{\circ x}$ is the solution of expected utility maximization problem \eqref{opt:eu} without risk constraint.
It is well known that $X_{\circ x}=(U^\prime)^{-1}(\lambda_{\circ}\rho)$ and hence that
$$\gamma_2(x)={1\over a}\log\Ept\left[e^{-a (U^\prime)^{-1}(\lambda_{\circ}\rho)}\right],$$ where $\lambda_{\circ}>0$ is the solution of equation
$$\Ept\left[\rho (U^\prime)^{-1}(\lambda\rho)\right]=x.$$

\begin{figure}[!ht]
\centering
\includegraphics[scale=0.30]{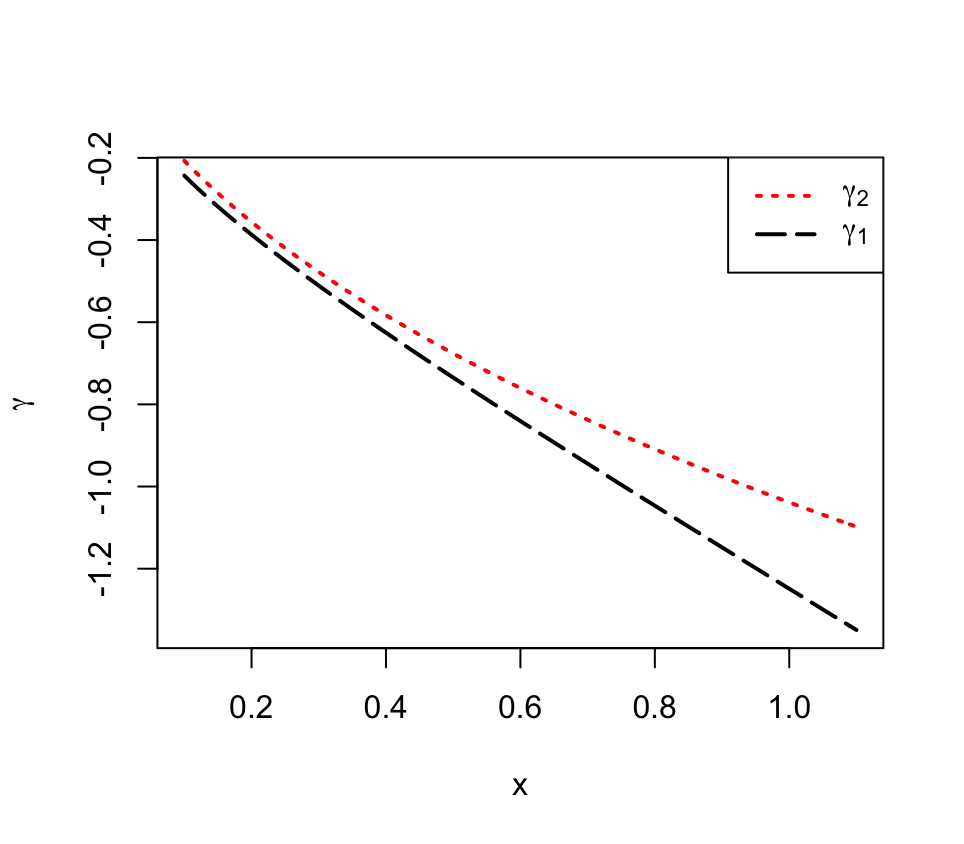}
\caption{\small
$\gamma_1$ and $\gamma_2$ for entropic risk measure when $a=2$, $\log\rho\sim\Nc(-0.5,1)$, and $U(x)=\log x$.}
\label{fig:gamma12_entropy}
\end{figure}

\begin{example}\label{exm:gamma2}
Assume that $U(x)=\log x$. In this case, $U^\prime(x)={1\over x}$ and $X_{\circ x}={1\over\lambda_{\circ}\rho}$. Therefore, $x=\Ept[\rho X_{\circ x}]={1\over\lambda_\circ}$ and $X_{\circ x}={x\over\rho}$. So we have $\gamma_2(x)={1\over a}\log\Ept[e^{-{ax\over\rho}}]$. 
Figure \ref{fig:gamma12_entropy} displays $\gamma_2$ when 
$\log\rho\sim\Nc(-0.5,1)$ and $a=2$. Particularly, $\gamma_2(1)=-1.0386$.
\end{example}

Let $x>0$ and $\gamma\in(\gamma_1(x),\gamma_2(x))$ be fixed. 
Now we assume that $X^*$ solves the expected utility maximization problem \eqref{opt:eu_X}. By Theorem \ref{thm:eu:X} and Lemma \ref{lma:duality:mu:gamma}, 
$$\Ept[e^{-aX^*}]=e^{ah(X^*)}=e^{a\gamma}.$$
Therefore, 
$$\psi(y)={e^{-ay}\over \Ept[e^{-aX^*}]}=e^{-a(y+\gamma)},\quad y>0$$
and 
$$\phi(y)=U^\prime(y)+\mu^*e^{-a(y+\gamma)},\quad y>0.$$
In the above, the Lagrangian multipliers $\lambda^*>0$ and $\mu^*>0$ are determined by
$$\Ept[e^{-aX^*}]=e^{a\gamma}\quad\mbox{and}\quad\Ept[\rho X^*]=x.$$

\newpage

\begin{figure}[!ht]
\centering
\includegraphics[scale=0.30]{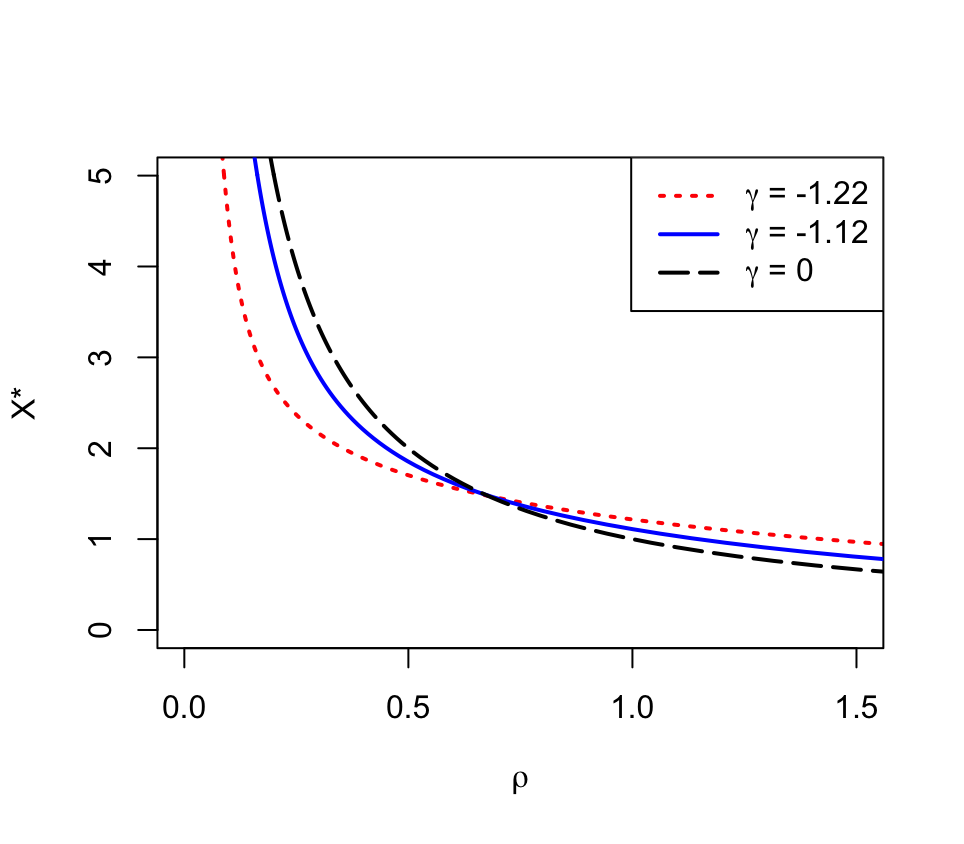}
\caption{\small
Expected utility maximizers $X^*$ v.s. $\rho$ for entropic risk measure, where $\gamma=0$ corresponds to the case without risk constraint.}
\label{fig:eu_entropy_xstar_rho}
\end{figure}

By \eqref{eq:integral:varphi:phi},  $\Ept[e^{-aX^*}]=e^{a\gamma}$ is equivalent to
\begin{equation}\label{eq:mu:eu;entropy}
1-a\displaystyle\int_0^\infty e^{-a y} F_\rho\left({\phi(y)\over\lambda^*}\right)dy= e^{a\gamma}.
 \end{equation}
By  \eqref{eq:U:psi:X}--\eqref{eq:pdf:X},
$$\Ept[\rho X^*]={1\over\lambda^*}\Ept[\phi(X^*)X^*]$$
and
$$F_{X^*}(y)=\Pbb(\lambda^*\rho\ge\phi(y))=1-F_\rho\left({\phi(y)\over\lambda^*}-\right),\quad y>0.$$
Then $\Ept[\rho X^*]=x$ is equivalent to
\begin{equation}\label{eq:lambda:eu:entropy}
-{1\over\lambda^*}\int_0^\infty y\phi(y)dF_\rho\left({\phi(y)\over\lambda^*}-\right)=x.
\end{equation}

Once $\lambda^*$ and $\mu^*$ have been determined by \eqref{eq:mu:eu;entropy}--\eqref{eq:lambda:eu:entropy}, then 
by \eqref{eq:U:psi:X}, 
$$X^*=\phi^{-1}(\lambda^*\rho).$$
So we also get the closed-form solution for the expected utility maximization problem \eqref{opt:eu_X} in the case when $h$ is an entropic risk measure. 

\begin{example} Assume that $\log\rho\sim\Nc(-0.5,1)$, $a=2$, and $U(x)=\log x$. When the budget level $x=1$, recall that $\gamma_1(1)=-1.2489$ and $\gamma_2(1)=-1.0386$.
In Figure \ref{fig:eu_entropy_xstar_rho}, the expected utility maximizers $X^*$ are displayed as functions of $\rho$ for risk levels $\gamma=-1.22$ and $-1.12$, as well as for $\gamma=0$ (corresponding to the case without risk constraint). The lower the risk level $\gamma$, i.e., the less the risk tolerance, the less risky is  the terminal payoff $X^*$ in the sense of He, Kouwenberg, and Zhou (2017, Definition 3.9 and Proposition 3.11):  the optimal terminal payoffs for two risk levels, as decreasing functions of the SDF $\rho$, have the single-crossing property. 
\end{example}

\subsection{Weighted Entropic Risk Measure}

For the optimization problems with risk controlled by a general WERM, the closed-form solutions are not available.
We can find the numerical solutions by the iterative methods in Sections \ref{sec:iteration:rmp} and \ref{sec:iteration:eu} and provide some numerical results for a specific model where the stochastic discount factor $\rho$, the utility function $U$, and the weighting measure $m$ are given as follows.
\begin{itemize}
\item  $\rho$ is log-normally distributed: $\log\rho\sim\Nc(b,\sigma^2)$, where $b=-0.5$ and $\sigma=1$.
\item  $U(x)=\log x$, $x>0$.
\item  $[a_0,a_1]=[2,3]$ and $m$ is the probability measure of the uniform distribution on $\{2.0, 2.1, 2.2,\dots,3.0\}$. In this case, 
\begin{equation}\label{eq:h:example}
h(X)={1\over11}\sum_{i=0}^{10}{1\over 2+{i\over10}}\log\Ept[e^{-(2+{i\over10})X}].
\end{equation}
\end{itemize}

We begin with the risk minimization problem.

\begin{figure}[!ht]
\centering
\includegraphics[scale=0.30]{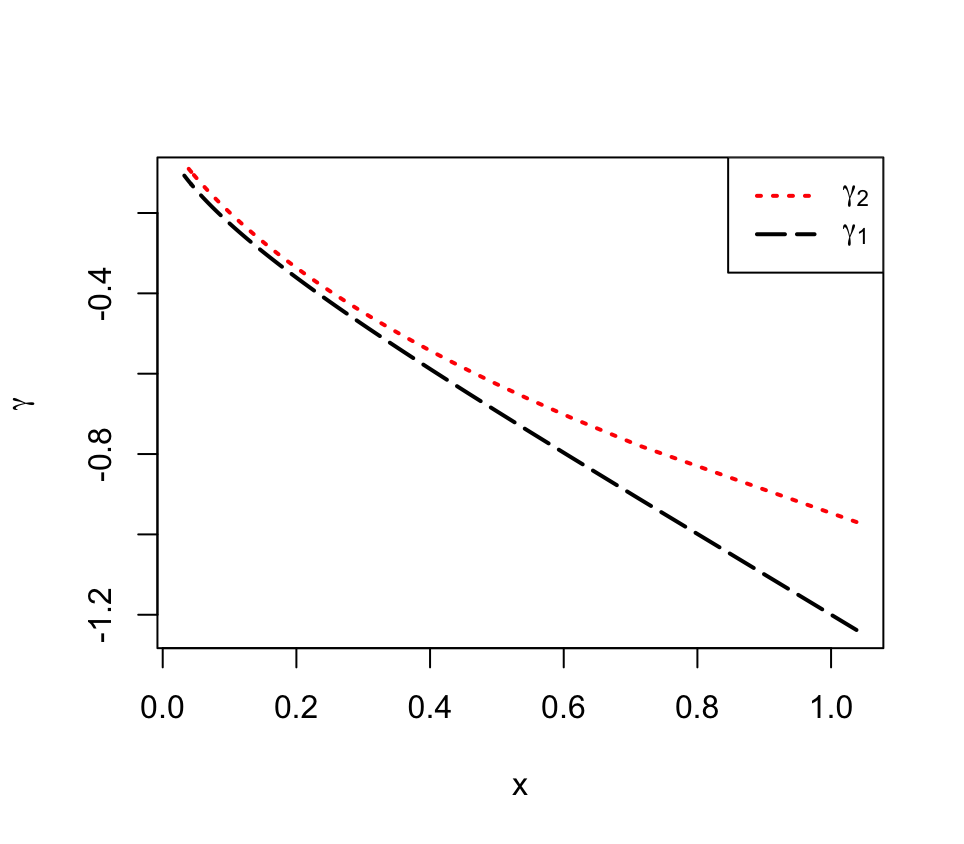}
\caption{\small
$\gamma_1$ and $\gamma_2$ for WERM \eqref{eq:h:example} when $\log\rho\sim\Nc(-0.5,1)$ and $U(x)=\log x$.}
\label{fig:gamma12_werm}
\end{figure}

 First, for some given $\lambda>0$, compute the optimal solution $X^*$ of problem \eqref{opt:rmp:mu} according to the iterative method in Section \ref{sec:iteration:rmp} and get the value of $\Xc(\lambda)$ according to  \eqref{eq:Xc:mu}. 
Obviously,  $\gamma_1(\Xc(\lambda))=h(X^*)$. By varying $\lambda$, we plot $\gamma_1$ v.s. $x$, as done in
Figure \ref{fig:gamma12_werm}. (By \eqref{eq:gamma2} and Example \ref{exm:gamma2}, $\gamma_2(x)=h\left({x/\rho}\right)$.) 

Second, for any given budget level $x>0$, say $x=1$, search for $\lambda^*>0$ such that $\Xc(\lambda^*)=x$ and then apply the iterative method to find the optimal solution of problem \eqref{opt:rmp} and get the value of $\gamma_1(1)$. For the specific model above, we get that $\gamma_1(1)=-1.2005$ and $\gamma_2(1)=-0.9489$.

Now we provide some numerical results for the expected utility maximization problem. 

First, for some given $\mu$ and $\lambda$, we can compute the optimal solution $X^{*\mu\lambda}$ of problem \eqref{opt:euwe:lm} according to the iterative method is Section \ref{sec:iteration:eu}  and get the value of $\Xc_\mu(\lambda)$ according to \eqref{eq:Xclambda}. Therefore, for the given $\mu$, we can search for $\lambda=\lambda(\mu)$ such that
$\Xc_\mu(\lambda(\mu))=x$. 

Second, for given $\mu$, we know that $X^{\mu\lambda(\mu)}$ is the solution of problem \eqref{opt:eu:mu0}, i.e.,
$X^{*\mu}=X^{*\mu\lambda(\mu)}$. Then we can compute $\Gamma(\mu)$ according to \eqref{eq:Gamma:mu}.

\begin{figure}[!ht]
\centering
\includegraphics[scale=0.30]{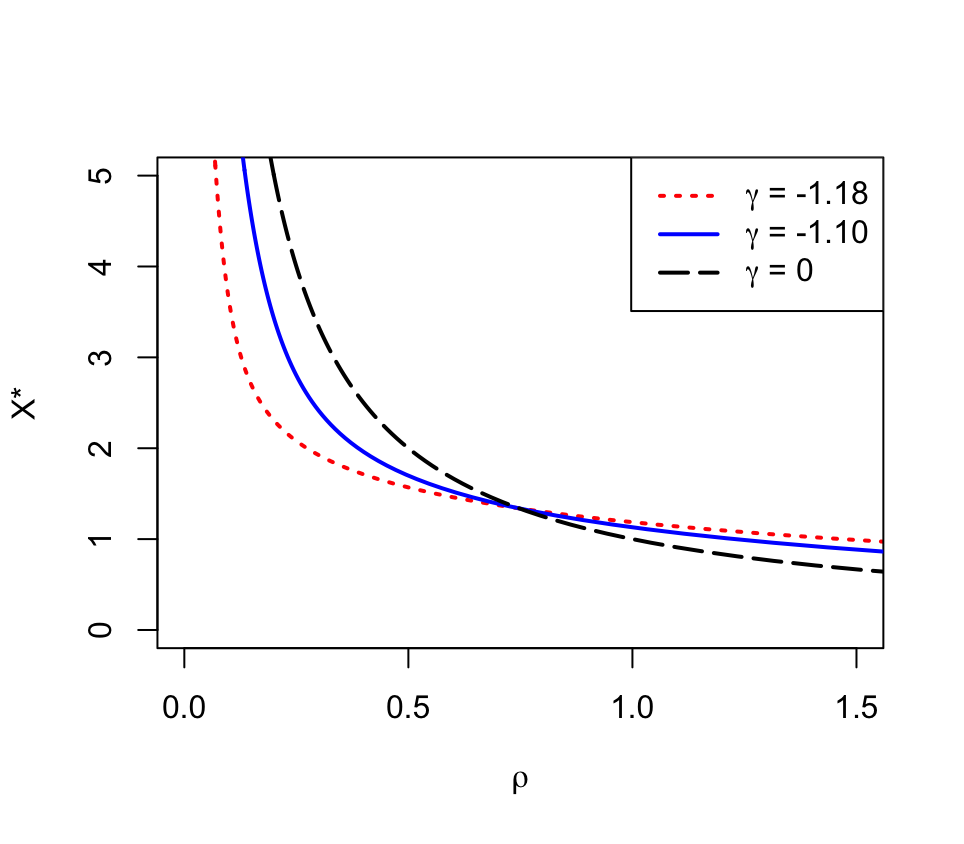}
\caption{\small
Expected utility maximizers $X^*$ v.s. $\rho$ for specific risk levels $\gamma$, where WERM is given by \eqref{eq:h:example} and $\gamma=0$ corresponds to the case without risk constraint}
\label{fig:eu_werm_xstar_rho}
\end{figure}

Third, search for $\mu^*$ such that $\Gamma(\mu^*)=\gamma$ and then apply the iterative method to find $X^{*\mu\lambda}$ for $\mu=\mu*$ and $\lambda=\lambda(\mu^*)$, which gives the optimal solution of problem  \eqref{opt:eu_X}. For the specific model above,  budget level $x=1$, and risk levels $\gamma=-1.18$, $-1.10$, and $0$,  the optimal solutions are displayed in Figure \ref{fig:eu_werm_xstar_rho}, where
$\gamma=0$ corresponding to the case without risk constraint. Similarly to Figure \ref{fig:eu_entropy_xstar_rho},
the optimal terminal payoffs for two risk levels, as decreasing functions of the SDF $\rho$, also have the single-crossing property.

\newpage

\appendix

\section*{Appendix}

\section{Some Properties of Weighted Entropic Risk Measures}\label{app:werm}

Recall that $h_a(X)$ is defined by \eqref{eq:ha:X} for all $a\in[0,\infty]$ and $X\in L^0$. A weighted entropic risk measure  $h: L^0\to[-\infty,\infty]$ is given by 
$$h(X)=\int_{[0,\infty]}h_a(X)m(da),\quad X\in L^0,$$
where the weighting measure $m$ is a probability measure on $[0,\infty]$.

\begin{definition}
For any $X_1, X_2\in L^0$, we say that $X_1-X_2$ is not constant if $\Pbb(X_1-X_2=c)<1$ for all $c\in\Rbb$.
\end{definition}

\begin{lemma}\label{lma:ha:strict:convex}
Let $a\in(0,\infty)$ and $X_1$, $X_2\in L^0$. If $X_1-X_2$ is not constant and both $h_a(X_1)$ and $h_a(X_2)$ are finite, then
$$h_a(\alpha X_1+(1-\alpha)X_2)< \alpha h_a(X_1)+(1-\alpha)h_a(X_2)\quad\mbox{for all }\alpha\in(0,1).$$
\end{lemma}

\proof 
For every $\alpha\in(0,1)$, we can find some $c_1, c_2\in\Rbb$ such that
$$\alpha c_1+(1-\alpha)c_2=0\mbox{ and } c_1-c_2=h_a(X_1)-h_a(X_2).$$ 
Then $h_a(X_1+c_1)=h_a(X_2+c_2)$, i.e., 
$$\beta\trieq\Ept[e^{-a(X_1+c_1)}]=\Ept[e^{-a(X_2+c_2)}].$$
By assumption, $\Pbb(X_1+c_1=X_2+c_2)<1$. Then the strict convexity of function $e^x$ implies that 
\begin{eqnarray*}
\Ept[e^{-a(\alpha X_1+(1-\alpha)X_2)}]&=&\Ept[e^{-a(\alpha (X_1+c_1)+(1-\alpha)(X_2+c_2))}]\\
&<&\alpha\Ept[e^{-a(X_1+c_1)}]+(1-\alpha)\Ept[e^{-a(X_2+c_2)}]=\beta.
\end{eqnarray*}
Therefore,
\begin{eqnarray*}
h_a(\alpha X_1+(1-\alpha)X_2)&<&{1\over a}\log \beta\\
&=&\alpha{1\over a}\log \Ept[e^{-a(X_1+c_1)}]+(1-\alpha){1\over a}\log \Ept[e^{-a(X_2+c_2)}]\\
&=&\alpha h_a(X_1+c_1)+(1-\alpha)h_a(X_2+c_2)\\
&=&\alpha h_a(X_1)+(1-\alpha)h_a(X_2)-\alpha c_1-(1-\alpha)c_2\\
&=&\alpha h_a(X_1)+(1-\alpha)h_a(X_2).
\end{eqnarray*}
\qed

\begin{theorem}\label{thm:h:strict:convex}
Assume $m((0,\infty))>0$.
Let $X_1$, $X_2\in L^0$. If $X_1-X_2$ is not constant and both $h(X_1)$ and $h(X_2)$ are finite, then
$$h(\alpha X_1+(1-\alpha)X_2)< \alpha h(X_1)+(1-\alpha)h(X_2)\quad\mbox{for all }\alpha\in(0,1).$$
\end{theorem}
\proof It is an abvious consequence of Lemma \ref{lma:ha:strict:convex}.\qed

The next lemma is obvious.

\begin{lemma}\label{lma:hahb} If $X\in L^0_+$, then $h_a(X)\le h_b(X)$ for all $0\le a<b\le\infty$. If $X\in L^0_+$ is not constant, then $h_a(X)<h_b(X)$ for all $0<a<b<\infty$.
\end{lemma}

\begin{lemma}\label{lma:liminf:ha}
Let $a\in(0,\infty)$ and a sequence $\{X_n, n\ge1\}\subset L^0_+$ converge to some $X\in L^0_+$ in probability. Then the sequence $\{h_a(X_n), n\ge1\}$ is bounded and non-positive. 
\end{lemma}

\proof It suffices to show that the sequence $\{h_a(X_n), n\ge1\}$ is lower bounded since $h_a(X)\le0$ for $a>0$ and $X\in L^0_+$. Suppose on the contrary that
 there exists some subsequence, which is still denoted by $\{h_a(X_n), n\ge1\}$, such that
 $\lim_{n\to\infty}h_a(X_n)=-\infty$, i.e., $\lim_{n\to\infty}\Ept[e^{-aX_n}]=0$. In this case, we have that $e^{-aX_n}\to0$ in probability, i.e.,  $X_n\to\infty$ in probability. This contradicts with the assumption that $X_n\to X\in L^0$ in probability. Therefore, $\{h_a(X_n), n\ge1\}$ is lower bounded.  \qed
 
 \begin{theorem}\label{thm:h:continuity}
 Let $m$ satisfy Assumption \ref{ass:m} and a sequence $\{X_n, n\ge1\}\subset L^0_+$ converge to some $X\in L^0_+$ in probability. Then $\lim_{n\to\infty}h(X_n)=h(X)$.
 \end{theorem}

\proof  The bounded convergence theorem implies that
$$h_a(X_n)={1\over a}\log \Ept[e^{-aX_n}]\to{1\over a}\log \Ept[e^{-aX}]=h_a(X)\quad\mbox{for all }a\in[a_0,a_1].$$
Lemmas  \ref{lma:hahb}--\ref{lma:liminf:ha} imply that
$$-\infty<\inf_{n\ge1}h_{a_0}(X_n)\le\sup_{n\ge1}h_{a_1}(X_n)\le0.$$
Lemma \ref{lma:hahb} implies that
$$h_{a_0}(X_n)\le h_a(X_n)\le h_{a_1}(X_n)\quad\mbox{for all }a\in [a_0,a_1]\mbox{ and }n\ge1.$$
Then by the bounded convergence theorem,
$$h(X_n)=\int_{[a_0,a_1]}h_a(X^n) m(da)\to\int_{[a_0,a_1]}h_a(X) m(da)=h(X).$$
\qed

\begin{theorem}\label{thm:h:direct:derivative}
Let $m$ satisfy Assumption \ref{ass:m}.
For any $X\in L^0_+$ and $Y\in L^1$ with $X+Y\in L^0_+$, we have that 
$$\lim_{t\downarrow0}{h(X+tY)-h(X)\over t}=\Ept[h^\prime(X)Y],$$
where $h^\prime(X)$ is given by \eqref{eq:hprime}.
\end{theorem}

\proof By the mean value theorem, there exists some measurable functions $$\alpha_1:[a_0,a_1]\times(0,1)\to[0,1]$$ 
and $$\alpha_2:[a_0,a_1]\times(0,1)\times\Omega\to[0,1]$$
such that, for all $(a,t)\in[a_0,a_1]\times(0,1)$,
\begin{align*}
\log \Ept[e^{-a(X+tY)}]-\log\Ept[e^{-aX}]
={\Ept[e^{-a(X+tY)}]-\Ept[e^{-aX}]\over
\alpha_1(a,t)\Ept[e^{-a(X+tY)}]+(1-\alpha_1(a,t))\Ept[e^{-aX}]}
\end{align*}
and
$$e^{-a(X+tY)}-e^{-aX}=-atYe^{-a(X+\alpha_2(a,t)tY)}\quad\mbox{a.s.}$$
Then for all $t\in(0,1)$, 
\begin{align*}
{h(X+tY)-h(X)\over t}=&\int_{[a_0,a_1]}{1\over ta}\left(\log\Ept[e^{-a(X+tY)}]-\log\Ept[e^{-aX}]\right)m(da)\\
=&\int_{[a_0,a_1]}{1\over ta}{\Ept[e^{-a(X+tY)}]-\Ept[e^{-aX}]\over
\alpha_1(a,t)\Ept[e^{-a(X+tY)}]+(1-\alpha_1(a,t))\Ept[e^{-aX}]}\, m(da)\\
=&-\int_{[a_0,a_1]}{\Ept\left[e^{-a(X+\alpha_2(a,t)tY)}Y\right]\over
\alpha_1(a,t)\Ept[e^{-a(X+tY)}]+(1-\alpha_1(a,t))\Ept[e^{-aX}]}\, m(da).
\end{align*}
Obviously, function $(a,t)\mapsto \Ept[e^{-a(X+tY)}]$ is strictly positive and continuous on $[a_0,a_1]\times[0,1]$ and hence there exists some $\varepsilon_0>0$ such that
$$\Ept[e^{-a(X+tY)}]>\varepsilon_0\quad\mbox{for all }(a,t)\in[a_0,a_1]\times[0,1].$$
Then by the dominated convergence theorem, we have that
$$\lim_{t\downarrow0}{h(X+tY)-h(X)\over t}=-\int_{[a_0,a_1]}{\Ept[e^{-aX}Y]\over \Ept[e^{-aX}]}\,m(da)=\Ept[h^\prime(X)Y].$$
\qed

\begin{theorem}\label{thm:h:ineq}
Let $m$ satisfy Assumption \ref{ass:m}. If $X,Y\in L^0_+$, then
$$h(Y)-h(X)\ge \Ept[h^\prime(X)(Y-X)].$$
\end{theorem}

\proof Assume that $X,Y\in L^0_+$. We first show that  
\begin{equation}\label{ineq:hh:hprime:0}
Y-X\in L^1\quad\Rightarrow\quad h(Y)-h(X)\ge \Ept[h^\prime(X)(Y-X)].
\end{equation}
Assume that $Y-X\in L^1$. For any $t\in[0,1]$, let $f(t)=h(X+t(Y-X))$. Then $f$ is convex on $[0,1]$ and hence
\begin{align*}
h(Y)-h(X)=f(1)-f(0)\ge\lim_{t\downarrow0}{f(t)-f(0)\over t}=\Ept[h^\prime(X)(Y-X)],
\end{align*}
where the last equality follows from Theorem \ref{thm:h:direct:derivative}. 
If $Y-X\notin L^1$, for any $k,n\ge1$, let
$$Z_{k,n}=\begin{cases}
 -k     & \text{if }Y-X\le-k, \\
 Y-X     & \text{if }-k<Y-X<n,\\
 n&\text{if }Y-X\ge n.
\end{cases}
$$
Obviously, $Z_{k,n}\in L^\infty$ for all $k,n\ge1$.  Moreover, 
$$Z_{k,n}=\min\{Y-X, n\}\ge0\quad\mbox{on}\quad (Y-X\ge0)$$
 and 
 $$Z_{k,n}=\max\{-k, Y-X\}\ge Y-X\quad \mbox{on} \quad (Y-X<0),$$ 
 yielding 
 $X+Z_{k,n}\in L^0_+$.
Then by \eqref{ineq:hh:hprime:0}, we have that
$$h(X+Z_{k,n})-h(X)\ge\Ept[h^\prime(X)Z_{k,n}],\quad k,n\ge1.$$
Noting that $h^\prime(X)\le0$ a.s., $h^\prime(X)\in L^\infty$ and $Z_{k,n}\le n$, we have that 
$$h^\prime(X)Z_{k,n}\ge n h^\prime(X)\in L^\infty.$$ Then applying Fatou lemma yields that
$$\liminf_{k\to\infty}\Ept[h^\prime(X)Z_{k,n}]\ge\Ept[h^\prime(X)Z_n],\quad n\ge1.$$
Here $Z_n=\min\{Y-X,n\}$, $n\ge1$. Moreover, Theorem \ref{thm:h:continuity} implies that
$$\lim_{k\to\infty}h(X+Z_{k,n})=h(X+Z_n).$$
Therefore, we have that
$$h(X+Z_n)-h(X)\ge\Ept[h^\prime(X)Z_n],\quad n\ge1.$$
Obviously, $h^\prime(X)Z_n\ge h^\prime(X)(Y-X)$ a.s. Then by Theorem \ref{thm:h:continuity} once again, we get that
$$h(Y)-h(X)=\lim_{n\to\infty}h(X+Z_n)-h(X)\ge\lim_{n\to\infty}\Ept[h^\prime(X)Z_n]\ge \Ept[h^\prime(X)(Y-X)].$$
\qed

\section{Proofs}\label{sec:proofs}

\subsection{Proof of Theorem \ref{thm:rmp:unique}}\label{sec: proof: thm:rmp:unique}
Let $x>0$.
Then there exists a sequence $\{X_n, n\ge1\}\subset \Xsc(x)$ such that
$h(X_n)\searrow \gamma_1(x)$ as $n\to\infty$. By Komol\'os lemma [see Kramkov and Schachermayer (1999, Lemma 3.1) for example], there exits a sequence
$\{X^{(n)}, n\ge1\}$ such that $X^{(n)}\in\mathrm{conv}(X_n, X_{n+1}, \dots)$ for all $n\ge1$ and $X^{(n)}$ converges in probability to some $[0,\infty]$-valued random variable $X^*$, where $\mathrm{conv}(\cdot)$ denotes the convex hull. Then by Fatou lemma, 
$$\Ept[\rho X^*]\le\liminf_{n\to\infty}\Ept[\rho X^{(n)}]\le x.$$
Therefore, $X^*\in \Xsc(x)$.  Then by \eqref{eq:hX:finite}, Theorem \ref{thm:h:continuity} and the convexity of $h$, 
$$-\infty<h(X^*)=\lim_{n\to\infty}h(X^{(n)})\le\gamma_1(x).$$
Therefore, $X^*$ solves problem \eqref{opt:rmp}.

Assume that  $X_1^*$ and $X_2^*$ solve problem \eqref{opt:rmp}. 
Obviously, we have that
$$\Ept[\rho X_1^*]=\Ept[\rho X_2^*]=x\mbox{ and }h(X_1^*)=h(X_2^*)=\gamma_1(x).$$
We need to show that $X_1^*=X_2^*$ a.s. 
Suppose on the contrary that $\Pbb(X_1^*=X_2^*)<1$. Then by $h(X_1^*)=h(X_2^*)=\gamma_1(x)$ we know that $X_1^*-X_2^*$ is not constant.  Therefore, by Theorem \ref{thm:h:strict:convex}, we have that
$$h(0.5 X_1^*+0.5X_2^*)<\gamma_1(x),$$  which is impossible since $0.5 X_1^*+0.5X_2^*\in\Xsc(x)$.

\subsection{Proof of Lemma \ref{lma:opt:rmp:mu0}} \label{app:proof:lma:opt:rmp:mu0}

The ``if" part is an implication of Luenberger (1969, Theorem 1 on p. 220). Now we show the ``only if" part.   Assume that $X^*$ solves problem \eqref{opt:rmp}.
Then by Luenberger (1969, Theorem 1 on p. 217), there exists some $\lambda^*\ge0$ such that
    \begin{align}\label{opt:rmp:lambda*0}
    \begin{cases}
      X^*\in\displaystyle\argmin_{X\in L^0_+} h(X)+\lambda^* \Ept[\rho X], \\
      \lambda^*(\Ept[\rho X^*]-x)=0.
\end{cases}
\end{align} 
It is easy to see that the optimization problem in \eqref{opt:rmp:lambda*0} has no solution if $\lambda^*=0$. Therefore, $\lambda^*>0$ and hence $\Ept[\rho X^*]=x$. 

Finally, by Luenberger (1969, Theorem 1 and the last paragraph on p. 222), we have $-\lambda^*\in\partial \gamma_1(x)$.

\subsection{Proof of Theorem \ref{thm:rmp}}\label{app:rmp}

Assume that $X^*$ solves problem \eqref{opt:rmp:mu}. Consider any fixed $X\in L^0_+$ with 
$X-X^*\in L^\infty$.
For any $\varepsilon\in(0,1)$, let
$X^\varepsilon=X^*+\varepsilon(X-X^*)$. 
Then by the optimality of $X^*$ and Theorem \ref{thm:h:direct:derivative}, 
\begin{eqnarray*}
0&\le& {1\over\varepsilon}\left(h(X^\varepsilon)+\lambda\Ept[\rho X^\varepsilon]-h(X^*)-\lambda\Ept[\rho X^*]\right)\\
&=&{1\over\varepsilon}\left( h(X^\varepsilon)- h(X^*)\right)+ \lambda\Ept[\rho(X-X^*)]\\
&\to&\Ept[(h^\prime(X^*)+\lambda\rho)(X-X^*)]\quad \text{as }\varepsilon\searrow0.
\end{eqnarray*}
Therefore, we have that 
\begin{equation*}
\begin{split}
&\Ept\left[(h^\prime(X^*)+\lambda\rho)(X-X^*)\right]\ge0\\
&\mbox{for all }X\in L^0_+\mbox{ with }X-X^*\in L^\infty,
\end{split}
\end{equation*}
which is equivalent to \eqref{eq:solution:rmp}.

Conversely, assume that $X^*$ satisfies condition \eqref{eq:solution:rmp}. Then by Theorem \ref{thm:h:ineq}, we have that 
\begin{eqnarray*}
h(X)+\lambda\Ept[\rho X]-h(X^*)-\lambda\Ept[\rho X^*]&\ge& \Ept[(h^\prime(X^*)+\lambda\rho)(X-X^*)]\\
&\ge&0 \quad\mbox{for all }X\in L^0_+.
\end{eqnarray*}
Therefore, $X^*$ solves problem \eqref{opt:rmp:mu}.
\qed

\subsection{Proof of Theorem \ref{thm:eu:X}}\label{app:proof:thm:eu:X*}

It is divided into a series of lemmas. 

\begin{lemma}\label{lma:X*>0}
Under Assumptions \ref{ass:U}--\ref{ass:m},  let $\lambda>0$ and $\mu\in(w_\lambda<\infty)$.
If $X^*$ solves problem \eqref{opt:euwe:lm}, then 
$X^*>0$ a.s. 
\end{lemma}

\proof 
Assume on the contrary that $\Pbb(X^*=0)>0$.
For any $\varepsilon\in(0,1)$, let $X^\varepsilon=X^*+\varepsilon\id_{\{X^*=0\}}$.
The concavity of $U$ and $U^\prime(0)=\infty$ imply that
\begin{eqnarray*}
{1\over \varepsilon}(\Ept[U(X^\varepsilon)]-\Ept[U(X^*)])&\ge&{1\over \varepsilon}\Ept[U^\prime(X^\varepsilon)(X^\varepsilon-X^*)]\\
&=&\Ept[U^\prime(X^\varepsilon)\id_{\{X^*=0\}}]\\
&=&U^\prime(\varepsilon)\Pbb(X^*=0)\nearrow\infty\quad\mbox{as }\varepsilon\searrow0.
\end{eqnarray*}
On the other hand,
$${1\over \varepsilon}(\Ept[\rho X^\varepsilon]-\Ept[\rho X^*])=\Ept[\rho\id_{\{X^*=0\}}]$$
and
$$\lim_{\varepsilon\downarrow0}{1\over \varepsilon}(h(X^\varepsilon)-h(X^*))=\Ept[h^\prime(X^*)\id_{\{X^*=0\}}]$$
by Theorem \ref{thm:h:direct:derivative}. 
Then we have that
$${1\over\varepsilon}\big(\left\{\Ept[U(X^\varepsilon)]-\lambda\Ept[\rho X^\varepsilon]-\mu h(X^\varepsilon)\right\}
-\left\{\Ept[U(X^*)]-\lambda\Ept[\rho X^*]-\mu h(X^*)\right\}\big)\to\infty$$
as $\varepsilon\searrow0$. This contradicts the optimality of $X^*$.
Therefore, $\Pbb(X^*=0)=0$ and hence $X^*>0$ a.s.\qed

\begin{lemma}\label{lma:Up2:X*}
Under Assumptions \ref{ass:U}--\ref{ass:m},  let $\lambda>0$ and $\mu\in(w_\lambda<\infty)$.
If $X^*$ solves problem \eqref{opt:euwe:lm}, then $U^\prime(X^*)-\lambda\rho-\mu h^\prime(X^*)\le0$ a.s. 
\end{lemma}

\proof Let $A=(U^\prime(X^*)-\lambda\rho-\mu h^\prime(X^*)>0)$.
 Suppose on the contrary that $\Pbb(A)>0$. Then $\Pbb(A_n)>0$ for some $n\ge1$, where
 $$A_n=\left(U^\prime(X^*)-\lambda\rho-\mu h^\prime(X^*)>{1\over n}\right).$$
For every $\varepsilon>0$, we can find a random variable $\xi\in(0,\varepsilon)$ a.s. such that
$$U^\prime(X^*+\xi)-\lambda\rho-\mu h^\prime(X^*)\ge {1\over 2n} \quad\mbox{a.s. on }A_n.$$ 
Let $X=X^*+\xi\id_{A_n}$. Then 
\begin{eqnarray*}
\int_{[a_0,a_1]}{e^{-aX}\over \Ept[e^{-aX}]}m(da)\ge\int_{[a_0,a_1]}{e^{-a(X^*+\xi\id_{A_n})}\over \Ept[e^{-aX^*}]}m(da)\ge e^{-a_1\varepsilon}\int_{[a_0,a_1]}{e^{-aX^*}\over \Ept[e^{-aX^*}]}m(da),
\end{eqnarray*}
implying that
$$-\mu h^\prime(X)\ge-\mu h^\prime(X^*)e^{-a_1\varepsilon}\quad\mbox{a.s.}$$
Observing that $-h^\prime(X^*)\in L^\infty_+$, we can let $\varepsilon$ (and $\xi$) be small enough such that
$$-\mu h^\prime(X)+\mu h^\prime(X^*)\ge(1-e^{-a_1\varepsilon})\mu h^\prime(X^*)\ge -{1\over 4n}\quad\mbox{a.s.}$$
By the concavity of $U$ and Theorem \ref{thm:h:ineq}, we have that
\begin{eqnarray*}
&&\{\Ept[U(X)]-\lambda\Ept[\rho X]-\mu h(X)\}-\{\Ept[U(X^*)]-\lambda\Ept[\rho X^*]-\mu h(X^*)\\
&\ge&\Ept[U^\prime(X)(X-X^*)]-\lambda\Ept[\rho(X-X^*)]+\mu\Ept[h^\prime(X)(X^*-X)]\\
&=&\Ept[(U^\prime(X^*+\xi)-\lambda\rho-\mu h^\prime(X))\xi\id_{A_n}]\\
&\ge&\Ept\left[\left(U^\prime(X^*+\xi)-\lambda\rho-\mu h^\prime(X^*)-{1\over 4n}\right)\xi\id_{A_n}\right]\\
&\ge&{1\over 4n}\Ept[\xi\id_{A_n}]>0,
\end{eqnarray*}
which contradicts the optimality of $X^*$.
\qed

\begin{lemma}\label{lma:Up:X*}
Under Assumptions \ref{ass:U}--\ref{ass:m},  let $\lambda>0$ and $\mu\in(w_\lambda<\infty)$.
If $X^*$ solves problem \eqref{opt:euwe:lm}, then $U^\prime(X^*)-\lambda\rho-\mu h^\prime(X^*)\ge0$ a.s. 
\end{lemma}

\proof Let $B=(U^\prime(X^*)-\lambda\rho-\mu h^\prime(X^*)<0)$.
 Suppose on the contrary that $\Pbb(B)>0$. Then $\Pbb(B_n)>0$ for some $n\ge1$, where
 $$B_n=\left(U^\prime(X^*)-\lambda\rho-\mu h^\prime(X^*)<-{1\over n}\right).$$
 By Lemma \ref{lma:X*>0}, $X^*>0$ a.s. 
For every $\varepsilon>0$, we can find a random variable $\xi\in(0,\varepsilon)$ a.s. such that
$X^*-\xi>0$ a.s. on $B_n$ and 
$$U^\prime(X^*-\xi)-\lambda\rho-\mu h^\prime(X^*)\le -{1\over 2n} \quad\mbox{a.s. on }B_n.$$ 
Let $X=X^*-\xi\id_{B_n}$. Then 
\begin{eqnarray*}
\int_{[a_0,a_1]}{e^{-aX}\over \Ept[e^{-aX}]}m(da)\le\int_{[a_0,a_1]}{e^{-a(X^*-\xi\id_{B_n})}\over \Ept[e^{-aX^*}]}m(da)\le e^{a_1\varepsilon}\int_{[a_0,a_1]}{e^{-aX^*}\over \Ept[e^{-aX^*}]}m(da),
\end{eqnarray*}
implying that
$$\mu h^\prime(X)\ge\mu h^\prime(X^*)e^{a_1\varepsilon}\quad\mbox{a.s.}$$
Observing that $-h^\prime(X^*)\in L^\infty_+$, we can let $\varepsilon$ be small enough such that
$$\mu h^\prime(X)-\mu h^\prime(X^*)\ge(e^{a_1\varepsilon}-1)\mu h^\prime(X^*)\ge -{1\over 4n}\quad\mbox{a.s.}$$
By the concavity of $U$ and Theorem \ref{thm:h:ineq}, we have that
\begin{eqnarray*}
&&\{\Ept[U(X)]-\lambda\Ept[\rho X]-\mu h(X)\}-\{\Ept[U(X^*)]-\lambda\Ept[\rho X^*]-\mu h(X^*)\\
&\ge&\Ept[U^\prime(X)(X-X^*)]-\lambda\Ept[\rho(X-X^*)]+\mu\Ept[h^\prime(X)(X^*-X)]\\
&=&\Ept[(-U^\prime(X^*-\xi)+\lambda\rho+\mu h^\prime(X))\xi\id_{B_n}]\\
&\ge&\Ept\left[\left(-U^\prime(X^*-\xi)+\lambda\rho+\mu h^\prime(X^*)-{1\over 4n}\right)\xi\id_{B_n}\right]\\
&\ge&{1\over 4n}\Ept[\xi\id_{B_n}]>0,
\end{eqnarray*}
which contradicts the optimality of $X^*$.
\qed

\begin{lemma}\label{lma:eu:XX*:L1}
Under Assumptions \ref{ass:U}--\ref{ass:m}, let $X^*\in L^0_+$. If $X^*$ satisfies condition \eqref{eq:eu:opt:con}, then $X^*$ solves problem \eqref{opt:euwe:lm}. 
\end{lemma}

\proof Let $X, X^*\in L^0_+$.  Assume that $X^*$ satisfies condition \eqref{eq:eu:opt:con}.
By the concavity of $U$,
$$\Ept[U(X)]-\Ept[U(X^*)]\le \Ept[U^\prime(X^*)(X-X^*)].$$
By Theorem \ref{thm:h:ineq},
$$h(X)-h(X^*)\ge \Ept[h^\prime(X^*)(X-X^*)].$$
Therefore,  condition \eqref{eq:eu:opt:con} implies that
\begin{align*}
&\big(\{\Ept[U(X)]-\lambda\Ept[\rho X]-\mu h(X)\}\big)-\{\Ept[U(X^*)]-\lambda\Ept[\rho X^*]-\mu h(X^*)\}\\
\le&\Ept[(U^\prime(X^*)-\lambda\rho-\mu h^\prime(X^*))(X-X^*)]=0.
\end{align*}
Thus, $X^*$ solves problem \eqref{opt:euwe:lm}.
\qed

\newpage

\section*{References}{
\begin{description}
\itemsep=0pt
\parskip=0pt

\item Acerbi, C. (2002): ``Spectral measures of risk: A coherent representation of subjective risk aversion,"
\textit{Journal of Banking and Finance} \textbf{26}, 1505--1518.

\item Acerbi, C., and P. Simonetti (2002): ``Portfolio optimization with spectral measures of risk," Unpublished manuscript.


\item Adam, A., M. Houkari and J.-P. Laurent (2008): ``Spectral risk measures and portfolio selection," \textit{Journal of Banking
and Finance} \textbf{32}, 1870--1882.


\item Basak, S., and A. Shapiro (2001): ``Value-at-risk-based risk
management: Optimal policies and asset prices," \textit{Review of
Financial Studies} \textbf{14}, 371--405.

\item Basak, S., A. Shapiro, and L. Tepl\'a (2006): ``Risk management
with benchmarking," \textit{Management Science} \textbf{52},
542--557.


\item Cahuich, L. D., and D. Hern\'andez-Hern\'andez (2013):
``Quantile Portfolio Optimization under Risk Measure Constraints,"
\textit{Appl. Math. Optim.} \textbf{68}, 157--179.


\item Ding, P., and Z. Q. Xu (2015): ``Rank-Dependent Utility
Maximization under Risk Exposure Constraint," working paper.

\item Feller, W. (1971): \textit{An Introduction to Probability Theory and its Applications, Vol. 2, (2nd Edition)}. New York: John Wiley \& Sons

\item F\"ollmer, H. and A. Schied (2016): \textit{Stochastic
Finance: An Introduction in Discrete Time (4th Edition)}. Berlin:
Walter de Gruyter.

\item Gerber, H. U. and Goovaerts, M. J. (1981): ``On the representation of additive principles of premium calculation,"
\textit{Scand. Actuarial J.} \textbf{4}, 221--227.

\item Goovaerts, M. J., Kaas, R., Laeven, R. J. A., Tang, Q. (2004): ``A comonotonic image of independence for additive risk measures," \textit{Insurance: Mathematics and Economics} \textbf{35}, 581--594.

\item He, X. D., H. Jin, and X. Y. Zhou (2015): ``Dynamic Portfolio Choice When Risk is Measured by Weighted VaR," \textit{Mathematics
of Operations Research} \textbf{40}: 773--796.

\item He, X. D., R. Kouwenberg, and X. Y. Zhou (2017): ``Rank-Dependent Utility and Risk Taking in Complete Markets," \textit{SIAM Journal on Financial Mathematics} \textbf{8}, 214--239.

\item Kramkov, D. and W. Schachermayer (1999): ``The Asymptotic Elasticity of Utility Functions and
Optimal Investment in Incomplete Markets," \textit{Ann. Appl. Prob.}
\textbf{9}, 904--950.

\item Luenberger, D. (1969): \textit{Optimization by Vector Space Methods}. New York: John Wiley \& Sons

\item Markowitz, H. (1952): ``Portfolio Selection," \textit{J. Finance} \textbf{7}, 77--91.

\item Mu, X.,  L. Pomatto, P. Strack, and O. Tamuz (2021): ``Monotone Additive Statistics," arXiv: 2102.00618.


\item Rockafellar, R. T., and S. Uryasev (2000): ``Optimization of
conditional value-at-risk," \textit{Journal of Risk} \textbf{2},
21--42.

\item Wang, X. and J. Xia (2021): ``Expected Utility Maximization with Stochastic Dominance Constraints in
Complete Markets," \textit{SIAM Journal on Financial Mathematics} \textbf{12}, 1054--1111.

\item Wei, P. (2018): ``Risk Management with Weighted VaR," \textit{Math.
Finance} \textbf{28}, 1020--1060.

\item Wei, P. (2021): ``Risk management with expected shortfall," \textit{Math Finan Econ}, forthcoming.

\end{description}
}

\end{document}